\documentclass{PoS}

\title{Studies of single top quark production at the Tevatron}

\ShortTitle{Studies of single top quark production at the Tevatron}

\author{\speaker{Breese Quinn}%
         \thanks{On behalf of the CDF and D0 Collaborations.}\\
        University of Mississippi\\
        E-mail: \email{quinn@phy.olemiss.edu}}

\abstract{In this paper we present several measurements of single top quark production from the 
CDF and D0 experiments at the Tevatron.  The various analyses utilize integrated luminosity ranging 
from 2.1 to 4.8 fb$^{-1}$.  The results include the observation of single top production with a combined 
cross section of $2.76^{+0.58}_{-0.47}$ pb for a top quark mass of 170 GeV/c$^2$, as well as measurements 
of top quark polarization and first evidence for $t$-channel production.}

\FullConference{35th International Conference of High Energy Physics\\
   July 22-28, 2010\\
    Paris, France}

\def\ttbar{${t\overline{t}}$ }

\begin{document}

\section{Introduction}
Within the Standard Model (SM), top quarks can be produced either as \ttbar pairs through the strong interaction, or 
singly in association with $b$ quarks through the electroweak interaction.  Studies of single top production rates and
properties could enhance understanding of SM processes, in particular yielding a direct measurement of the CKM 
parameter $\mid V_{tb}\mid$.  These measurements are also sensitive to the effects of several new physics scenarios.  
In proton-antiproton collisions at the Tevatron, the cross section for single top production is expected to be nearly 
half the cross section for \ttbar\cite{tt_pred,t_pred}.  However, even though strong production of top was discovered 
in1995\cite{tt_disc}, it took 12 more years to find evidence for the electroweak counterpart\cite{t_evid} due to much 
more challenging backgrounds. 

\begin{figure}
\begin{center}
\includegraphics[width=.5\textwidth]{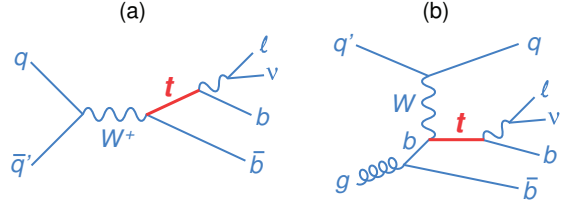}
\caption{Feynman diagrams for (a) $s$-channel and (b) $t$-channel single top quark production, with subsequent 
top decay to $b$ quark, charged lepton and neutrino.}
\label{feynman}
\end{center}
\end{figure}

In Tevatron collisions with $\sqrt{s} = 1.96$ TeV, $s$- and $t$-channel single top production modes dominate, with 
expected cross sections of $\sigma_{s} = 1.12\pm 0.05$ pb and $\sigma_{t} = 2.34\pm 0.13$ pb (at $m_t=170$ GeV/c$^2$).  
Figure~\ref{feynman} shows diagrams of these two processes, where the $W$ boson from the $t$ decay subsequently decays 
to a charged lepton and neutrino offering the cleanest signature for experimental observation of single top.  In this 
paper, several results from the CDF and D0 experiments examining these lepton+jets events in 2.1 to 4.8 fb$^{-1}$ of 
integrated luminosity are presented.

\section{Observation}

The main background obscuring single top signal is $W+$jets events (including light and heavy flavor jets) which have
a similar topology and cross section some four orders of magnitude greater.  Other major backgrouonds are \ttbar 
and multijet production, with smaller contributions from dibosons and $Z+$jets.  Monte Carlo simulations are used to 
model the signal and all backgrounds except multijets, which is obtained from data samples.  $W+$jets and 
multijets are normalized to the data, while all other backgrounds are normalized to SM NNLO cross section 
calculations.

Events selected from the data must include an isolated, high transverse momentum electron or muon, missing transverse 
energy ($\not E_T$) from the escaped neutrino, and associated $b$ jet all from the top decay.  $s$-channel events have
a second $b$ jet from the $b$ produced with the top, and $t$-channel events have an additional light quark jet and can 
have a second $b$ jet.  Therefore, selected events must have 2 or 3 (or 4 for D0) jets, 1 or 2 of which must be 
identified as $b$ jets ($b$-tagged).  After selection, signal to background is only on the order of 1:20, so 
multivariate techniques are employed to further isolate the signal.

Both experiments perform multiple multivariate anaylses (MVAs) in order to maximize the discriminating power of many
separate variables.  CDF and D0 each perform boosted decision tree, neural net, and matrix element analyses, and CDF 
employs a likelihood function and $s$-channel only likelihood function.  CDF also performs a neural net analysis on an
orthogonal $\not E_T$+jets sample to recover events with taus from the $W$ decay\cite{met_cdf}.  For each MVA, the data
is separated into  several individual analysis channels based on number of jets, number of $b$-tags, and for D0 the 
lepton type.  The cross section is determined using a likelihood formed as a product over all channels. Finally, an 
overall cross section measurement for each experiment is made by combining the separate MVAs.  D0 does this with a 
bayesian neural net, CDF with a neural net for lepton+jets followed by a simultaneous fit of lepton+jets and 
$\not E_T$+jets.  

\begin{figure}
\begin{center}
\includegraphics[width=.33\textwidth]{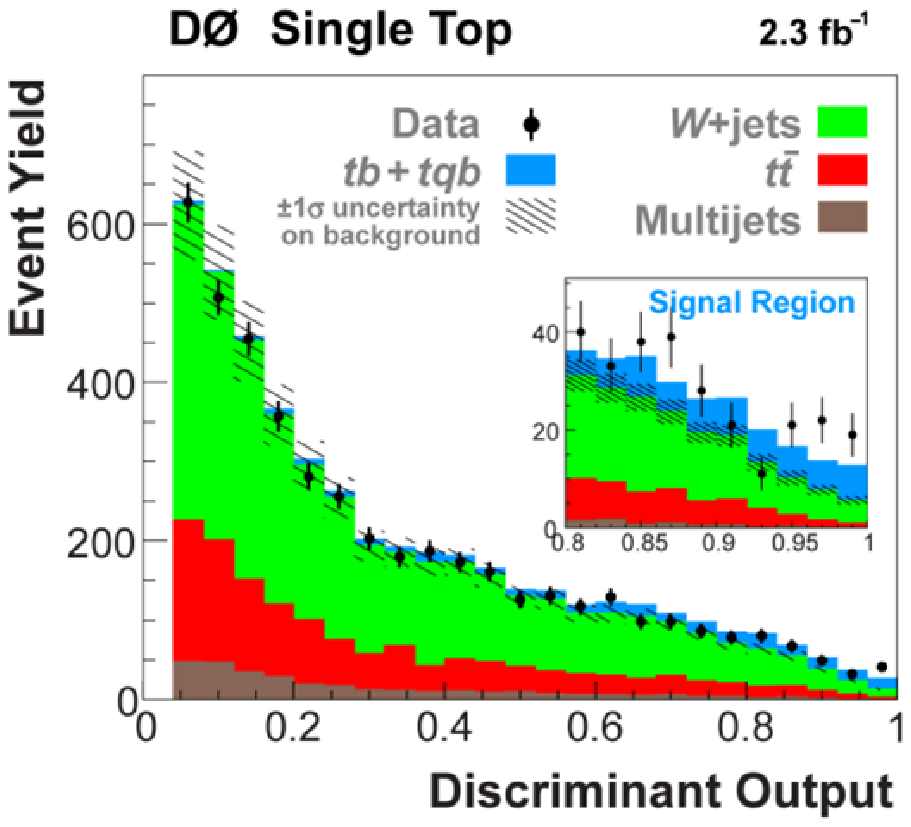}
\includegraphics[width=.4\textwidth]{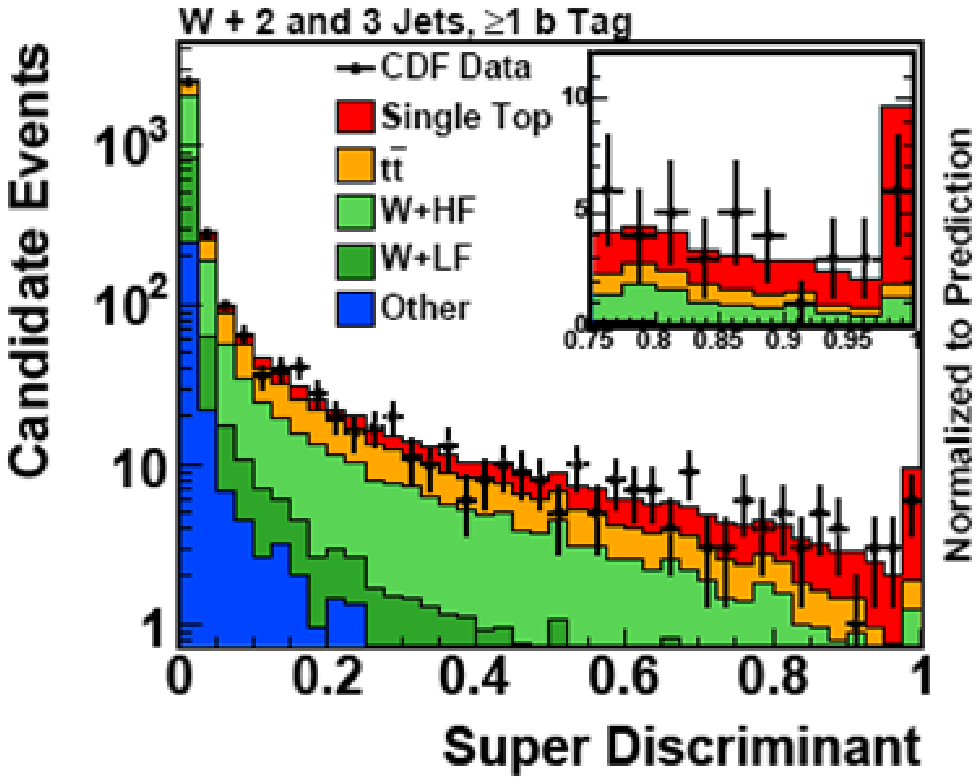}
\caption{Discriminants combining all lepton+jets multivariate analyses for D0 (left) and CDF (right).}
\label{D0_CDF}
\end{center}
\end{figure}

\begin{figure}
\begin{center}
\includegraphics[width=.3\textwidth]{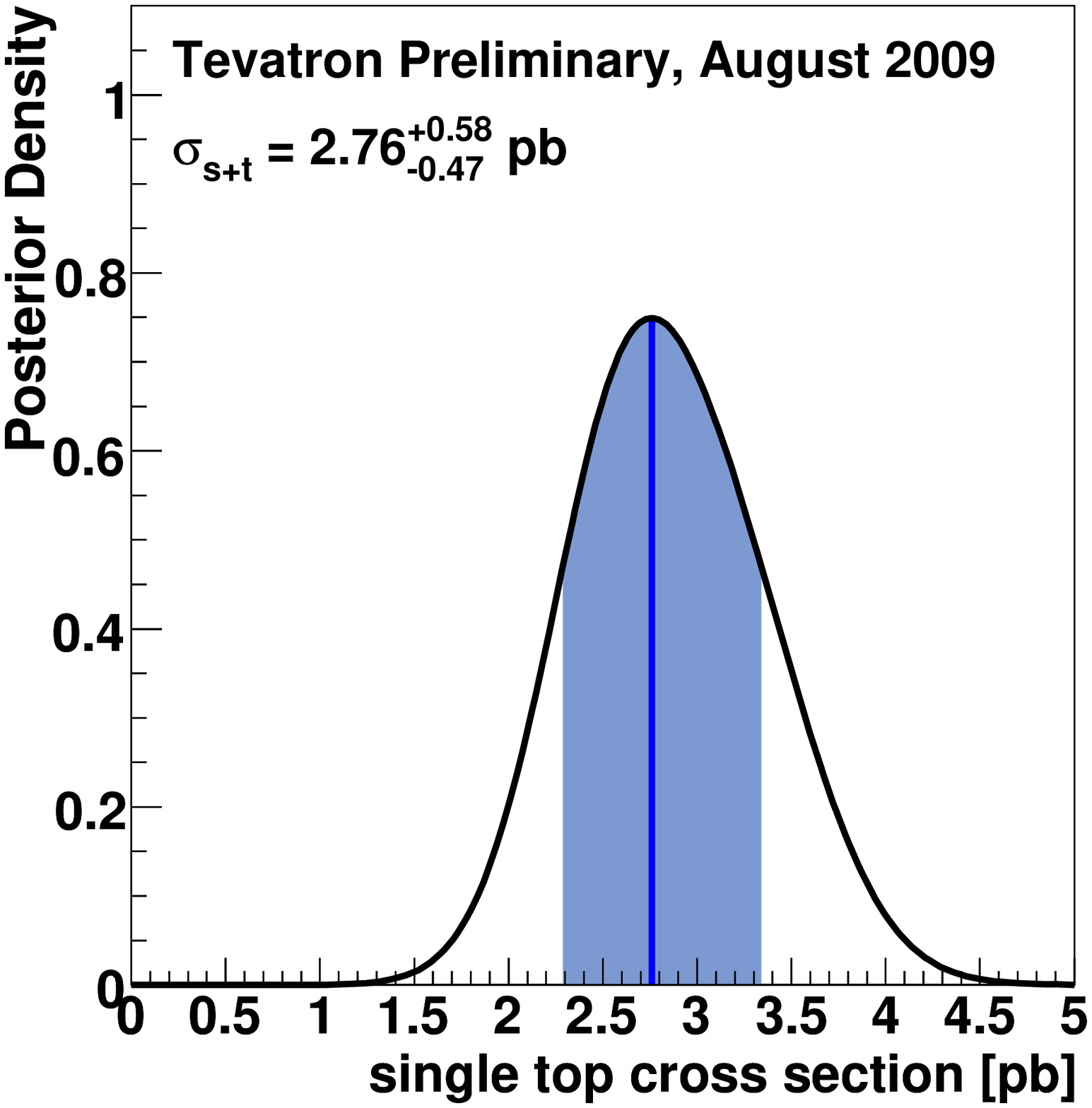}
\includegraphics[width=.3\textwidth]{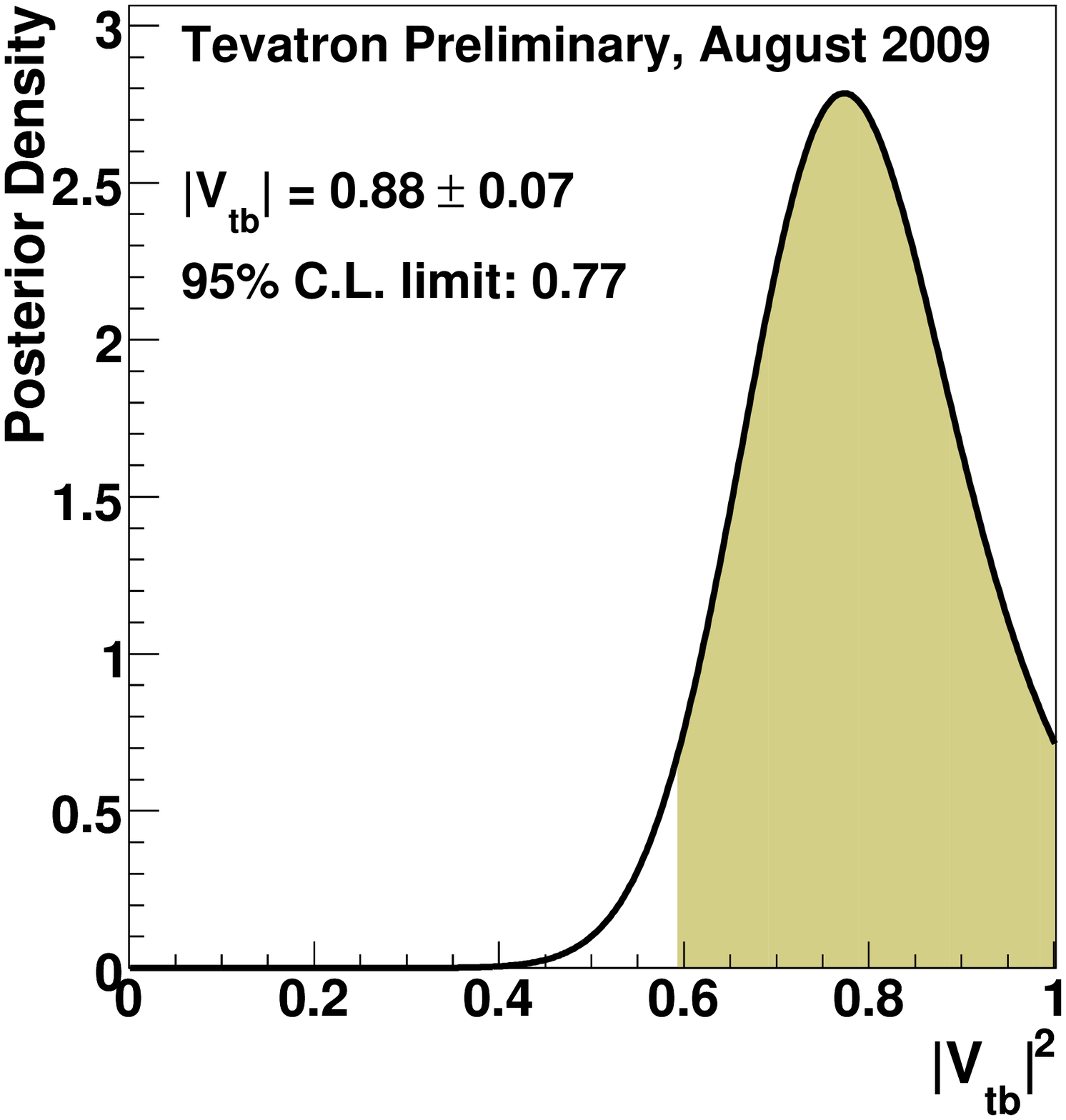}
\caption{Posterior probability distributions of the Tevatron combination for the single top production cross 
section for $m_t=170$ Gev/c$^2$ (left), and $\mid V_{tb}\mid$ assuming a theory cross section of 3.46 pb (right).}
\label{tev_post}
\end{center}
\end{figure}

\begin{figure}
\begin{center}
\includegraphics[width=.4\textwidth]{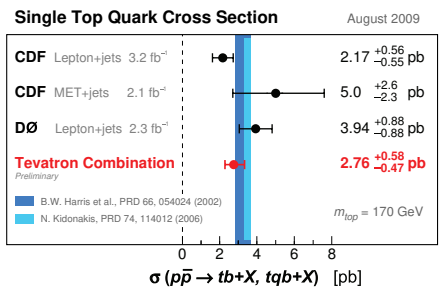}
\caption{Combination of CDF and D0 single top cross section measurements, with theoretical predictions.}
\label{tev_xsec}
\end{center}
\end{figure}

The combined lepton+jets discriminants are shown in Figure~\ref{D0_CDF}, which illustrates the signal separation key 
to the single top production observations by both experiments published on the same day\cite{t_obs}.  The combined 
$s$-channel and $t$-channel cross section measurements and extracted values and lower limits on $\mid V_{tb}\mid$ 
(assuming $0\leq\mid V_{tb}\mid^2\leq1$) are detailed in Table~\ref{obs}.  The CDF and D0 results were then combined 
with a Bayesian analysis using all nine CDF and D0 MVA discriminants as inputs, and the posterior probability 
densities for the cross section and $\mid V_{tb}\mid$ can be seen in Figure~\ref{tev_post}\cite{tev_comb}.  The 
Tevatron measurement of $\sigma_{s+t} = 2.76^{+0.58}_{-0.47}$ pb is consistent with the SM, and the two experiments' 
values are compatible with each other at the level of $1.6\sigma$ (Figure~\ref{tev_xsec}).       

\begin{table}
\begin{center}
\begin{tabular}{|c||c|c|c|c|c|c|}
\hline
    & Luminosity        & Cross Section          & Exp Sig      & Obs Sig      & $\mid V_{tb}\mid$ Meas & $\mid V_{tb}\mid$ Lower Lim \\ \hline\hline
D0  & 2.3 fb$^{-1}$     & $3.94\pm 0.88$ pb      & $4.5\sigma$  & $5.0\sigma$  & $1.07\pm 0.12$         & 0.78         \\ \hline
CDF & 2.1-3.2 fb$^{-1}$ & $2.3^{+0.6}_{-0.5}$ pb & $5.9\sigma$  & $5.0\sigma$  & $0.91\pm 0.13$         & 0.71         \\ \hline
\end{tabular}
\caption{Single top production cross section and $\mid V_{tb}\mid$ results for CDF and D0.}
\label{obs}
\end{center}
\end{table}

\section{Other Measurements}

\begin{figure}
\begin{center}
\includegraphics[width=.35\textwidth]{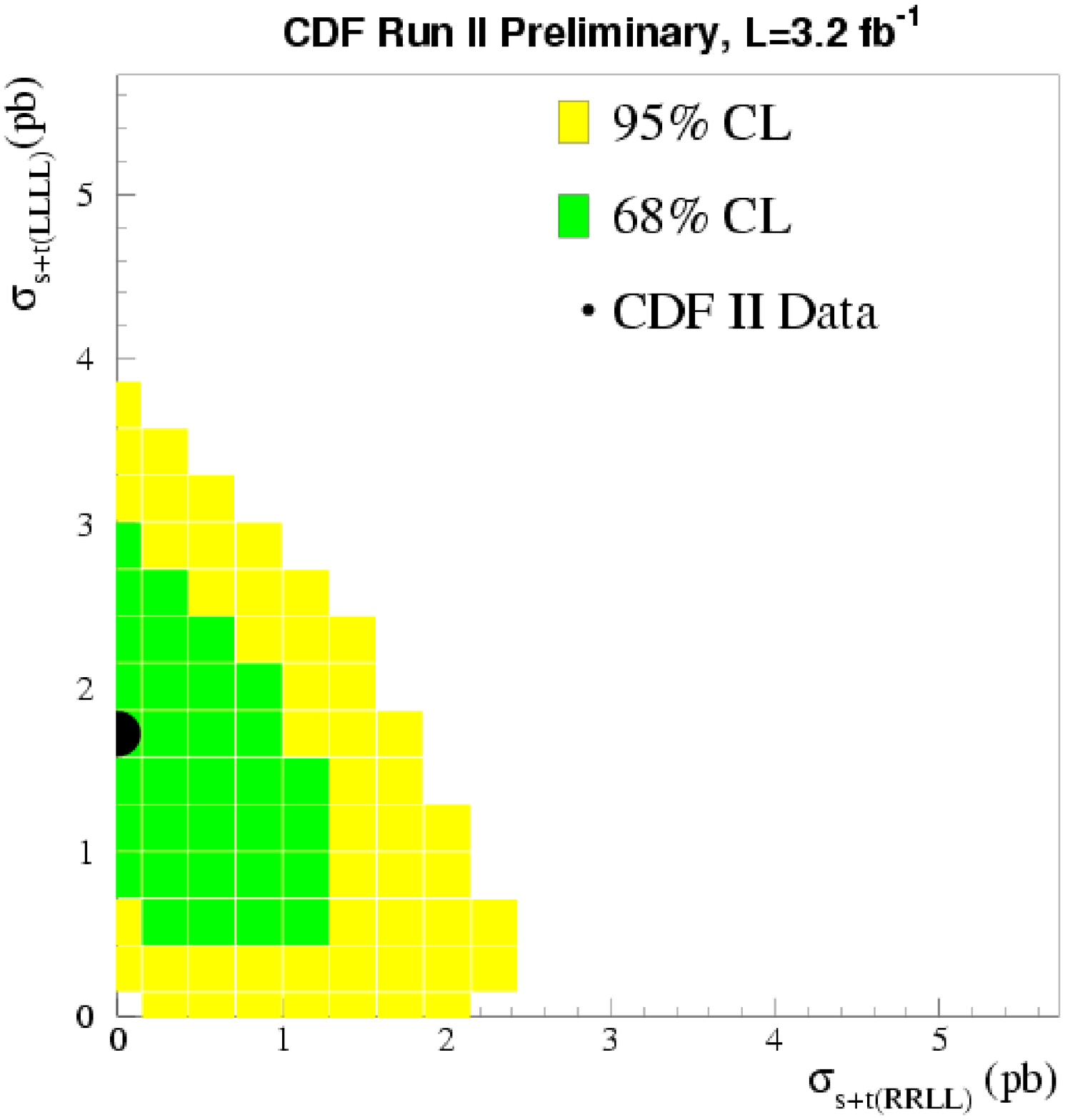}
\includegraphics[width=.35\textwidth]{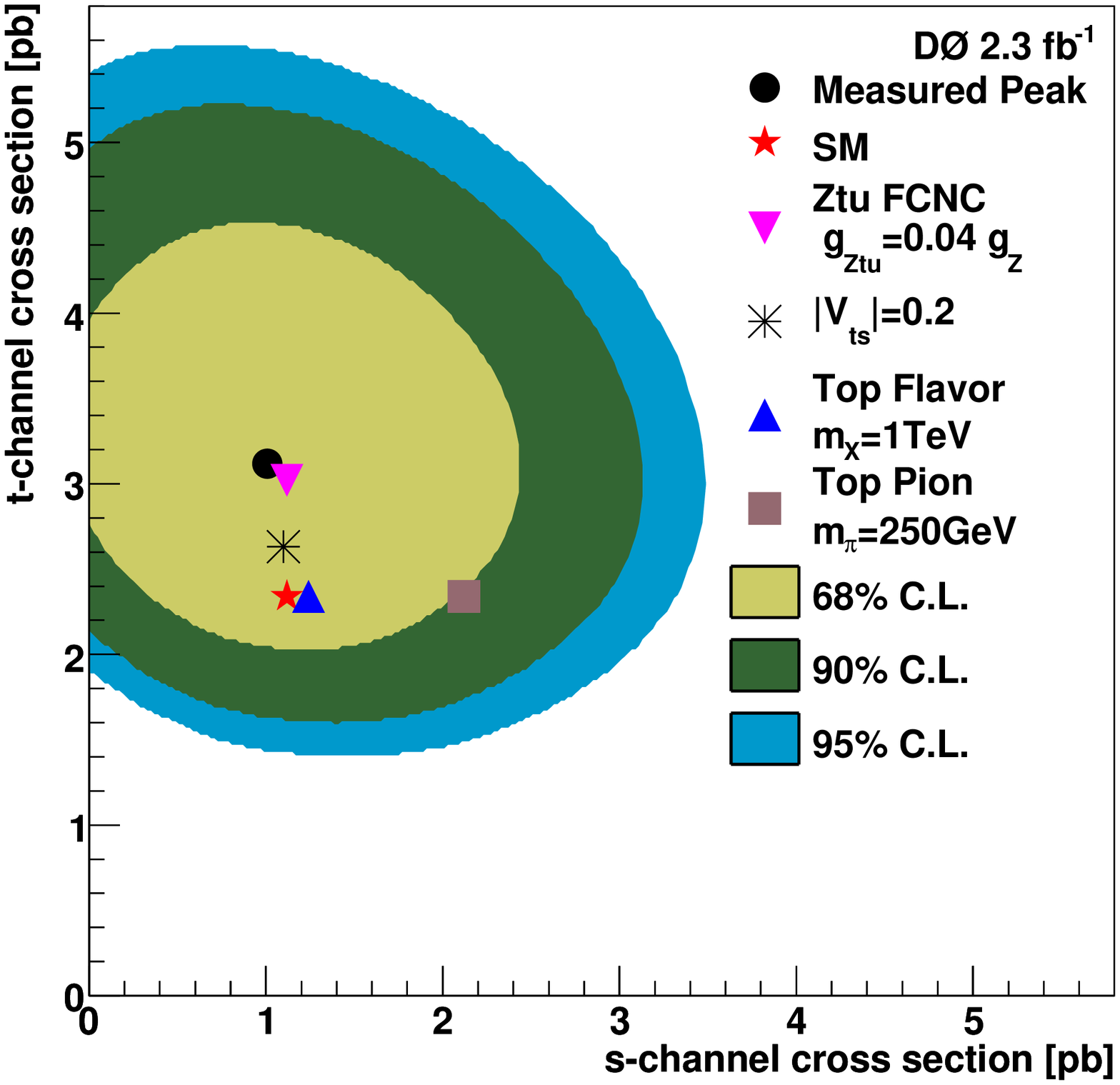}
\caption{2D fit of SM left-handed (LLLL) vs. anomalous right-handed (RRLL) production cross sections (left), and  
posterior probability of $t$-channel and $s$-channel production cross sections (right).}
\label{other}
\end{center}
\end{figure}

Several other studies of single top quarks have been completed since the joint discovery was made.  Within the SM,
single top quarks are produced purely through $V-A$ interactions, and thus should have 100\% left-handed 
polarization.  However, non-SM production can introduce right-handed $V+A$ couplings, so a polarization measurement 
different than pure left would be an indication of new physics.  CDF performed a 2D analysis with separate 
discriminants for left-handed production and decay (LLLL), and right-handed production with left-handed decay (RRLL) 
(Figure~\ref{other}, left).  With measured cross sections of $\sigma_{LLLL} = 1.72$ and $\sigma_{RRLL} = 0$, and 
polarization $\sigma_R-\sigma_L / \sigma_R+\sigma_L = -1^{+1.5}_{-0}$, no evidence for right-handed couplings was 
found. 

The $t$-channel production rate is particularly sensitive to flavor changing neutral currents and anomalous couplings, 
thus D0 retrained the MVAs to treat $s$-channel single top as part of the background and $t$-channel single top as the 
signal.  In this analysis, the individual $s$- and $t$-channel cross sections are measured simultaneously, without 
constraining $\sigma_t/\sigma_s$ to the SM value\cite{tchan_D0}.  The plot of $\sigma_t$ vs. $\sigma_s$ in 
Figure~\ref{other}, right shows a ratio consistent with the SM, and a measured value of $\sigma_t=3.1\pm{0.9}$, 
representing the first evidence for $t$-channel single top production with an obvserved significance of $4.8\sigma$.  
D0 has also published a single top cross section measurement using boosted decision trees to analyze tau+jets events, 
with a result of $\sigma_{s+t} = 3.4^{+2.0}_{-1.8}$ pb\cite{tau_D0}, and a bayesian neural net search which found no 
evidence for flavor changing neutral currents in single top production\cite{fcnc_D0}.

\section{Summary}

Several studies of single top quarks have been completed at the Tevatron, all employing a variety of multivariate
analysis techniques to  extract the signal from an extremely challenging background environment.  After the CDF and 
D0 discovery of single top quark production, the experiments' analyses have been combined to yield Tevatron 
measurements of $\sigma_{s+t} = 2.76^{+0.58}_{-0.47}$ pb and $\mid V_{tb}\mid > 0.77$.  Among other results are the
CDF measurement of left-handed top quark polarization, and D0's first evidence for $t$-channel production.  More 
single top results with greater precision will be forthcoming as the Tevatron data sets will more than double by 
2011.


\begin{thebibliography}{99}
  \bibitem{tt_pred} S. Moch and P. Uwer, \emph{Phys. Rev. D} {\bf 78}, 034003 (2008) [{\tt arXiv:0804.1476v2}].
  \bibitem{t_pred} N. Kidonakis, \emph{Phys. Rev. D} {\bf 74}, 114012 (2006) [{\tt arXiv:hep-ph/0609287v3}].
  \bibitem{tt_disc} F. Abe et al. (CDF Collaboration), \emph{Phys. Rev. Lett.} {\bf 74}, 2626 (1995) 
    [{\tt arXiv:hep-ex/9503002v2}]; S. Abachi et al. (D0 Collaboration), \emph{Phys. Rev. Lett.} {\bf 74}, 2632 (1995)
    [{\tt arXiv:hep-ex/9503003v1}].
  \bibitem{t_evid} V.M. Abazov et al. (D0 Collaboration), \emph{Phys. Rev. Lett.} {\bf 98}, 181802 (2007) 
    [{\tt arXiv:hep-ex/0612052v2}]; T. Aaltonen et al. (CDF Collaboration), \emph{Phys. Rev. Lett.} {\bf 101}, 252001 
    (2008) [{\tt arXiv:0809.2581v2}]; V.M. Abazov et al. (D0 Collaboration), \emph{Phys. Rev. D} {\bf 78}, 012005 
    (2008) [{\tt arXiv:0803.0739v2}].
  \bibitem{met_cdf} F. Abe et al. (CDF Collaboration), \emph{Phys. Rev. D} {\bf 81}, 072003 (2010) 
    [{\tt arXiv:1001.4577v1}].
  \bibitem{t_obs} V.M. Abazov et al. (D0 Collaboration), \emph{Phys. Rev. Lett.} {\bf 103}, 092001 (2009) 
    [{\tt arXiv:0903.0850v2}]; T. Aaltonen et al. (CDF Collaboration), \emph{Phys. Rev. Lett.} {\bf 103}, 
    092002 (2009) [{\tt arXiv:0903.0885v3}]; T. Aaltonen et al. (CDF Collaboration), \emph{Accepted by Phys. Rev. D} 
    [{\tt arXiv:1004.1181v3}].
  \bibitem{tev_comb} Tevatron Electroweak Working Group (CDF and D0 Collaborations), FERMILAB-TM-2440-E (2009) 
    [{\tt arXiv:0908.2171v1}]. 
  \bibitem{tchan_D0} V.M. Abazov et al. (D0 Collaboration), \emph{Phys. Lett. B} {\bf 682}, 363 (2010) 
    [{\tt arXiv:0907.4259v2}].
  \bibitem{tau_D0} V.M. Abazov et al. (D0 Collaboration), \emph{Phys. Lett. B} {\bf 690}, 5 (2010) 
    [{\tt arXiv:0912.1066v2}].
  \bibitem{fcnc_D0} V.M. Abazov et al. (D0 Collaboration), \emph{Phys. Lett. B} {\bf 693}, 81 (2010) 
    [{\tt arXiv:1006.3575v2}].

\end{thebibliography}
\end{document}